\documentclass[twocolumn]{aastex61}
\pdfoutput=1

\usepackage{xspace}

\newcommand{\project}[1]{\textsl{#1}}
\newcommand{\nustar}{\project{NuSTAR}\xspace}

\newcommand{\astrosat}{\project{Astrosat}\xspace}
\newcommand{\ixpe}{\project{IXPE}\xspace}

\newcommand{\sref}{Section~\ref}
\newcommand{\deadt}{\ensuremath{t_d}\xspace}

\newcommand{\rms}{\ensuremath{\mathrm{rms}}\xspace}

\newcommand{\Normal}{\ensuremath{{\mathcal N}}\xspace}

\received{XXXX}
\revised{XXXX}
\accepted{\today}
\submitjournal{ApJL}

\shorttitle{The Fourier Amplitude Difference correction for periodograms}
\shortauthors{Bachetti \& Huppenkothen}
\begin{document}

\title{No time for dead time - Use the Fourier amplitude differences to normalize dead time-affected periodograms}

\correspondingauthor{Matteo Bachetti}
\email{bachetti@oa-cagliari.inaf.it}

\author[0000-0002-4576-9337]{Matteo Bachetti}
\affil{INAF-Osservatorio Astronomico di Cagliari, via della Scienza 5, I-09047 Selargius (CA)}

\author{Daniela Huppenkothen}
\affiliation{Center for Data Science, New York University, 60 5h Avenue, 7th Floor, New York, NY 10003 }
\affiliation{Center for Cosmology and Particle Physics, Department of Physics, New York University, 4 Washington Place, New York, NY 10003, USA}
\affiliation{DIRAC Institute, Department of Astronomy, University of Washington, 3910 15th Ave NE, Seattle, WA 98195}

%% Note that the \and command from previous versions of AASTeX is now
%% depreciated in this version as it is no longer necessary. AASTeX 
%% automatically takes care of all commas and "and"s between authors names.

%% AASTeX 6.1 has the new \collaboration and \nocollaboration commands to
%% provide the collaboration status of a group of authors. These commands 
%% can be used either before or after the list of corresponding authors. The
%% argument for \collaboration is the collaboration identifier. Authors are
%% encouraged to surround collaboration identifiers with ()s. The 
%% \nocollaboration command takes no argument and exists to indicate that
%% the nearby authors are not part of surrounding collaborations.

%% Mark off the abstract in the ``abstract'' environment. 
\begin{abstract}
Dead time affects many of the instruments used in X-ray astronomy, by producing a strong distortion in power density spectra. This can make it difficult to model the aperiodic variability of the source or look for quasi-periodic oscillations.
Whereas in some instruments a simple a priori correction for dead-time-affected power spectra is possible, 
this is not the case for others such as \nustar, where the dead time is non-constant and long ($\sim$2.5\,ms).
\citet{Bachetti+15} suggested the cospectrum obtained from light curves of independent detectors within the same instrument as a possible way out, but this solution has always only been a partial one: the measured \rms was still affected by dead time, because the width of the power distribution of the cospectrum was modulated by dead time in a frequency-dependent way.

In this Letter, we suggest a new, powerful method to normalize dead-time-affected cospectra and power density spectra. Our approach uses the difference of the Fourier amplitudes from two independent detectors to characterize and filter out the effect of dead time. This method is crucially important for the accurate modeling of periodograms derived from instruments affected by dead time on board current missions like \nustar and \astrosat, but also future missions such as \ixpe.
\end{abstract}

%% Keywords should appear after the \end{abstract} command. 
%% See the online documentation for the full list of available subject
%% keywords and the rules for their use.
\keywords{X-rays: binaries --- 
X-rays: general --- methods: data analysis --- methods: statistical}

%% From the front matter, we move on to the body of the paper.
%% Sections are demarcated by \section and \subsection, respectively.
%% Observe the use of the LaTeX \label
%% command after the \subsection to give a symbolic KEY to the
%% subsection for cross-referencing in a \ref command.
%% You can use LaTeX's \ref and \label commands to keep track of
%% cross-references to sections, equations, tables, and figures.
%% That way, if you change the order of any elements, LaTeX will
%% automatically renumber them.

%% We recommend that authors also use the natbib \citep
%% and \citet commands to identify citations.  The citations are
%% tied to the reference list via symbolic KEYs. The KEY corresponds
%% to the KEY in the \bibitem in the reference list below. 

%%%%%%%%%%%%%%%%%%%%%%%% SECTION 1 %%%%%%%%%%%%%%%%%%%%%%%%%%

\section{Introduction} \label{sec:intro}
Dead time is an unavoidable and common issue of photon-counting instruments.
It is the time $\deadt$ that the instrument takes to process an event and be ready for the next event.
In most current astronomical photon-counting X-ray missions, dead time is of the \textit{non-paralyzable} kind, meaning that the instrument does not accept new events during dead time, avoiding a complete lock of the instrument if the incident rate of photons is higher than $1/\deadt$.
Being roughly energy-independent, dead time is not usually an issue for spectroscopy, as it only affects the maximum rate of photons that can be recorded, so it basically only increases the observing time needed for high quality spectra.

For timing analysis, the effect of dead time is far more problematic.
Dead time heavily distorts the periodogram, the most widely used statistical tool to estimate the power density spectrum (PDS)%
\footnote{here we will use the term PDS for the actual source power spectrum, and \textit{periodogram} to indicate our estimate of it, or otherwise said, the realization of the ``real'' power spectrum we observe in the data}%
, with a characteristic pattern that is stronger for brighter sources. 
It is often not possible to disentangle this power spectral distortion due to dead time and the broadband noise components characterizing the emission of accreting systems.
In the special case where dead time is constant, its shape can be modeled precisely \citep{Zhang+95,Vikhlinin+94}.
However, dead time is often different on an event-to-event basis (e.g., in \nustar), and it is not obvious how to model it precisely, also because the information on dead time is often incomplete in the data files distributed by HEASARC \citep[see, e.g.][]{Bachetti+15}.
%\footnote{Whereas in principle this information could be obtained by using the PRIOR column in the unfiltered event files for some missions, the live time given in this column is affected by events that are not recorded in the file, like shield vetos in the case of \nustar, and the estimate of dead time is necessarily uncertain}%

%For a more thorough discussion about different dead time behaviors see \citet{Zhang+95}.

When using data from missions carrying two or more \textit{identical and independent} detectors like \nustar, \citet{Bachetti+15} proposed an approach to mitigate instrumental effects like dead time exploiting this redundancy: 
where in standard analysis, light curves of multiple detectors are summed before Fourier transforming the summed light curve, 
it is possible to instead Fourier-transform the signal of two independent detectors and combine the Fourier amplitudes in a \textit{cospectrum} -- the real part of the cross spectrum -- instead of the periodogram. 
Since dead time is uncorrelated between the two detectors, the resulting powers have a mean white noise level fixed to 0, which resolves the first and most problematic issue created by dead time (see details in \citealt{Bachetti+15}); however, the resulting powers no longer follow the statistical distribution expected for power spectra, and their probability distribution is frequency-dependent.
Whereas a noise cospectrum in the absence of dead time would follow a Laplace distribution \citep{HuppenkothenBachetti18},
dead time affects the width of the probability distribution for cospectral powers and modulates the measured \rms similarly to the distortion acted on power spectra.
In this Letter, we show a method to precisely recover the shape of the periodogram by looking at the difference of the Fourier amplitudes of the light curves of two independent detectors.
This difference, in fact, contains information on the uncorrelated noise produced by dead time, but not on the source-related  signal which is correlated between the two detectors.
This allows to disentangle the effects of dead time from those of the source variability.

In \sref{sec:fourierdiff} we show that, in the absence of dead time, the difference of the Fourier amplitudes calculated from two independent detectors contains the sum of the correlated signal (the source signal) and uncorrelated noise (detector-related noise), and that their difference eliminates the source part. 
In \sref{sec:fadsec} we use extensive simulations to show how to use this fact to correct dead-time-affected periodograms, and we describe the limitations of this method.

%%%%%%%%%%%%%%%%%%%%%%%% SECTION 2 %%%%%%%%%%%%%%%%%%%%%%%%%%

\section{On the difference of Fourier amplitudes} \label{sec:fourierdiff}

\begin{figure*}
\plottwo{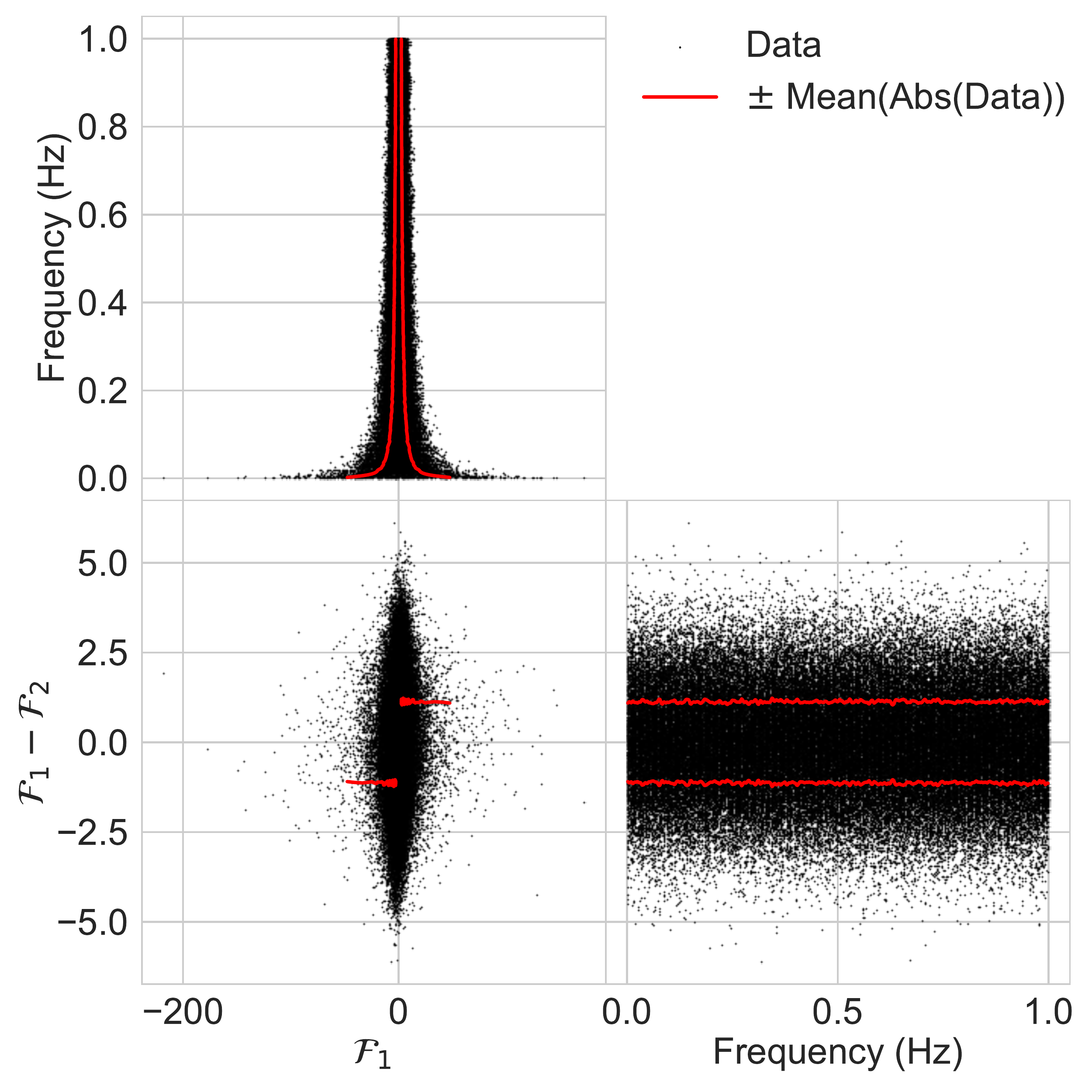}{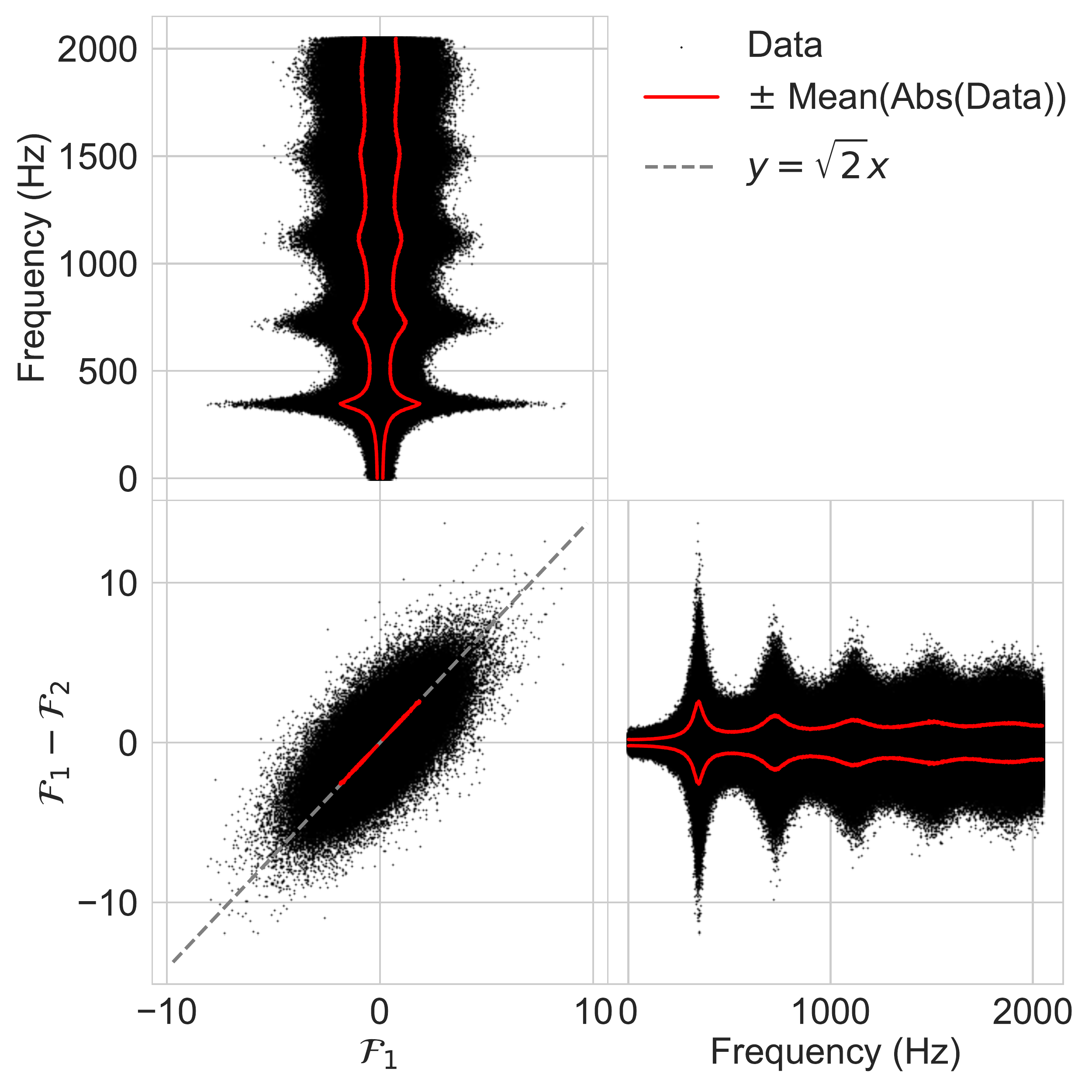}
\caption{Real-valued Fourier amplitudes obtained by single light curves ($\mathcal{F}_1$) and difference between two realizations of the same source light curve ($\mathcal{F}_1 - \mathcal{F}_2$), plotted against each other in two cases: (Left) Strong $1/f$ red noise and no dead time, calculated over many 500\,s segments of the light curve, and (Right) no red noise and strong dead time, calculated over 5\,s segments of the light curve. 
The red curve gives the frequency-dependent spread of the distributions, measured by the mean of the absolute values of the curves in each frequency bin. 
The different behavior of Fourier amplitude differences in the two cases is evident: in the dead-time-free case, the Fourier amplitude difference does \textit{not} correlate with the Fourier amplitude, while in the dead-time-affected case, this follows a precise linear relationship.
}
\label{fig:fourierdiff}
\end{figure*}

Let us consider two identical and independent detectors observing the same variable source, producing independent and strictly simultaneous time series, with identical even sampling $\delta t$, $\mathbf{x} = \{x_k\}_{k=1}^N$ and $\mathbf{y} = \{y_k\}_{k=1}^N$. For a stochastic process (e.g.\ $1/\nu$-type red noise), the Fourier amplitudes will vary as a function of $N_{\mathrm{phot}}P(\nu)/4$, where $P(\nu)$ (Leahy-normalized, \citealt{Leahy+83}) is the shape of the power spectrum underlying the stochastic process, and $N_{\mathrm{phot}}$ denotes the number of photons in a light curve. 

If the two detectors observe the same source simultaneously, the amplitudes and phases of the stochastic process will be shared among $\mathbf{x}$ and $\mathbf{y}$, while each light curve will be affected \textit{independently} by the photon counting noise in the detector, as well as the dead time process. 

Dead time can be considered a convolution on the signal \citep{Vikhlinin+94}. 
Following the convolution theorem the Fourier transform $\mathcal{F}$ of dead time-affected light curves will be the \textit{product} of the Fourier transform of the signal $\mathcal{S}$ and the Fourier transform of the dead time filter $\mathcal{D}$:
\begin{equation}
\mathcal{F}(\nu) = \mathcal{S}(\nu)\cdot\mathcal{D}(\nu)
\end{equation}
For a large enough number of data points $N$, the complex Fourier amplitudes $S_j$ will be composed of a sum of two independent random normal variables for the intrinsic red noise variability and the detector photon counting noise, respectively: $S_j = S_{sj} + S_{nj}$, with $\Re (S_{sj}) \sim \Normal(0, \sigma_{sj}^2)$ and $\Re(S_{nj}) \sim \Normal(0, \sigma_n^2)$, and similarly for the imaginary parts.
The red noise variance $\sigma_{sj}^2 = \sigma_{s}^2(\nu) = N_\mathrm{phot}P(\nu)/4$ (where $N_{\mathrm{phot}} = \sum_{k=1}^{N}{x_k}$) is given by the power spectrum of the underlying stochastic process and is frequency-dependent. However, the photon counting noise $\sigma_n^2 = N_\mathrm{phot}/2$ is independent of frequency. 
Note that $S_{xsj} = S_{ysj}$, because the amplitudes of the stationary noise process will be the same for the Fourier transforms of $\mathbf{x}$ and $\mathbf{y}$ for the case considered here, while the components due to white noise differ between the two time series.
The dead time filter affects the sum of signal and white noise amplitudes as a multiplicative factor and only depends on count rate, which is equal for both light curves given identical detectors.
Thus, the difference between the Fourier amplitudes for the two time series $\mathbf{x}$ and $\mathbf{y}$ will be:
\begin{equation}
F_{xj} - F_{yj} = (S_{xj} - S_{yj})\cdot D_j = (S_{xnj} - S_{ynj})\cdot D_j
\end{equation}
Because $S_{xsj} = S_{ysj}$, but $S_{xnj} \neq S_{ynj}$ (since the white noise component is formed in each detector separately), the \textit{difference} of the real and imaginary Fourier amplitudes between the two light curves effectively encodes the white noise component only, multiplied by the Fourier transform of the dead time filter.
This fact effectively allows us to separate out the (source-intrinsic) red noise from the spurious variability introduced by dead time: if we can extract the shape of the dead time filter $|\mathcal{D}|^2$ from the Fourier amplitude differences of the two detectors, we can use it to correct the shape of the periodogram. In the following section, we lay this procedure out in more detail, and describe its limits in Section~\ref{sec:caveat}.

%%%%%%%%%%%%%%%%%%%%%%%% SECTION 3 %%%%%%%%%%%%%%%%%%%%%%%%%%%%%%

\section{The FAD method} \label{sec:fadsec}
\subsection{Data simulation} \label{sec:data}
All simulated sets in this paper were produced and analyzed with a combination of the two Python libraries \texttt{stingray}%
\footnote{The library is under heavy development.
    For this work we used the version identified by the hash \texttt{3e64f3d}.
    See \href{https://github.com/StingraySoftware/notebooks/}{https://github.com/StingraySoftware/notebooks/} for tutorials on simulations, light curve production and timing analysis with Stingray}
\citep{huppenkothen2016} and \texttt{HENDRICS} v.\texttt{3.0b2} \citep[formerly known as MaLTPyNT;][]{2015ascl.soft02021B}, both affiliated to Astropy \citep{astropy2013,astropy18}.

We used the same procedure and algorithms described by \citet{Bachetti+15}, Section 4, which we briefly summarize here.
We used the \citet{timmer1995} method to create a red noise light curve starting from a given power spectral shape. 
This method is implemented in the \texttt{stingray.simulate} module.
This step needs to be done carefully: if the initial light curves contain significant random noise, the process for the creation of events creates a random variate on the top of the local count rate--which is varying randomly already--producing a non-Poissonian final light curve. 
We initially simulated light curves with a very high mean ``count rate'' such that the Poisson noise was relatively small. We then renormalized the light curves to the wanted (lower) count rate and \rms and finally used these light curves to simulate event lists using rejection sampling, implemented in the \texttt{stingray.Eventlist.simulate\_times()} method.
Then, the \texttt{hendrics.fake.filter\_for\_deadtime()} function was used to apply a non-paralyzable dead time of 2.5\,ms to the simulated event lists. For more details on the simulated data sets and the validation of the simulation infrastracture, see also Section \ref{sec:correction} and the available Jupyter notebooks%
\footnote{\href{https://github.com/matteobachetti/deadtime-paper-II/}{https://github.com/matteobachetti/deadtime-paper-II/}}%
\footnote{\href{https://github.com/StingraySoftware/HENDRICS/tree/master/notebooks}{https://github.com/StingraySoftware/HENDRICS/tree/master/notebooks}}.
After producing these synthetic event lists, we started the standard timing analysis: we produced light curves with a bin time of $\sim0.244$\,ms, and calculated power spectral products (cospectrum, periodogram) over segments of these light curves using \texttt{stingray}.

%%%%%%%%%%%%%%%%%%%%%%%% SECTION 4 %%%%%%%%%%%%%%%%%%%%%%%%%%

\subsection{First test: white noise} \label{sec:wndeadtime}

\begin{figure*}
\plotone{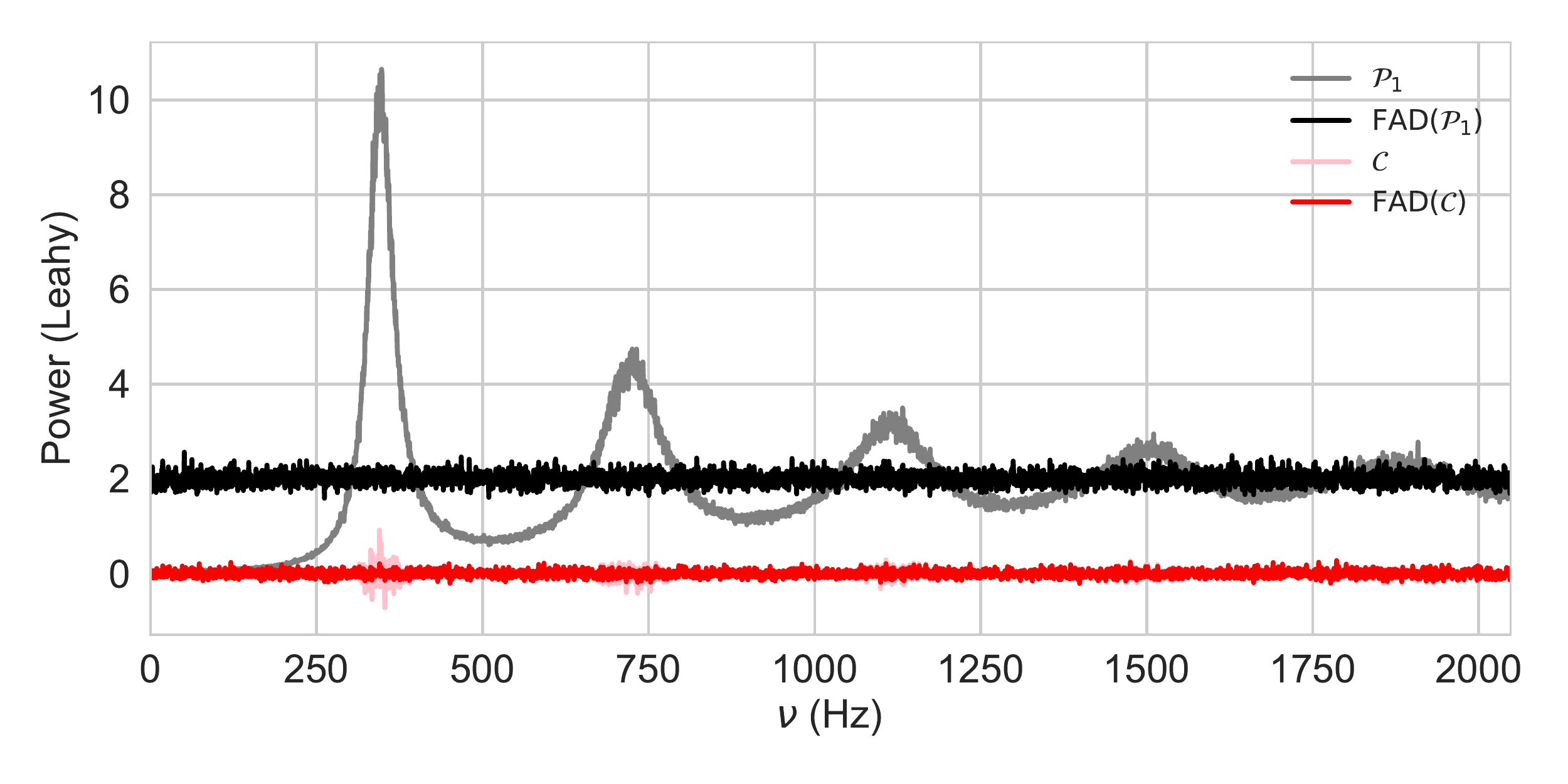}
\caption{Periodogram and cospectrum, before and after FAD correction, for a pure white noise light curve (count rate 2000 ct/s). 
The dead-time-driven distortion of the white noise level in the periodogram, and the frequency-dependent modulation of the \rms in both spectra, disappear after applying the FAD correction. 
We averaged 500 periodograms calculated over 2-sec intervals, to decrease the scatter and highlight the distortion of powers.}
\label{fig:comparison}
\end{figure*}

\begin{figure*}
\plotone{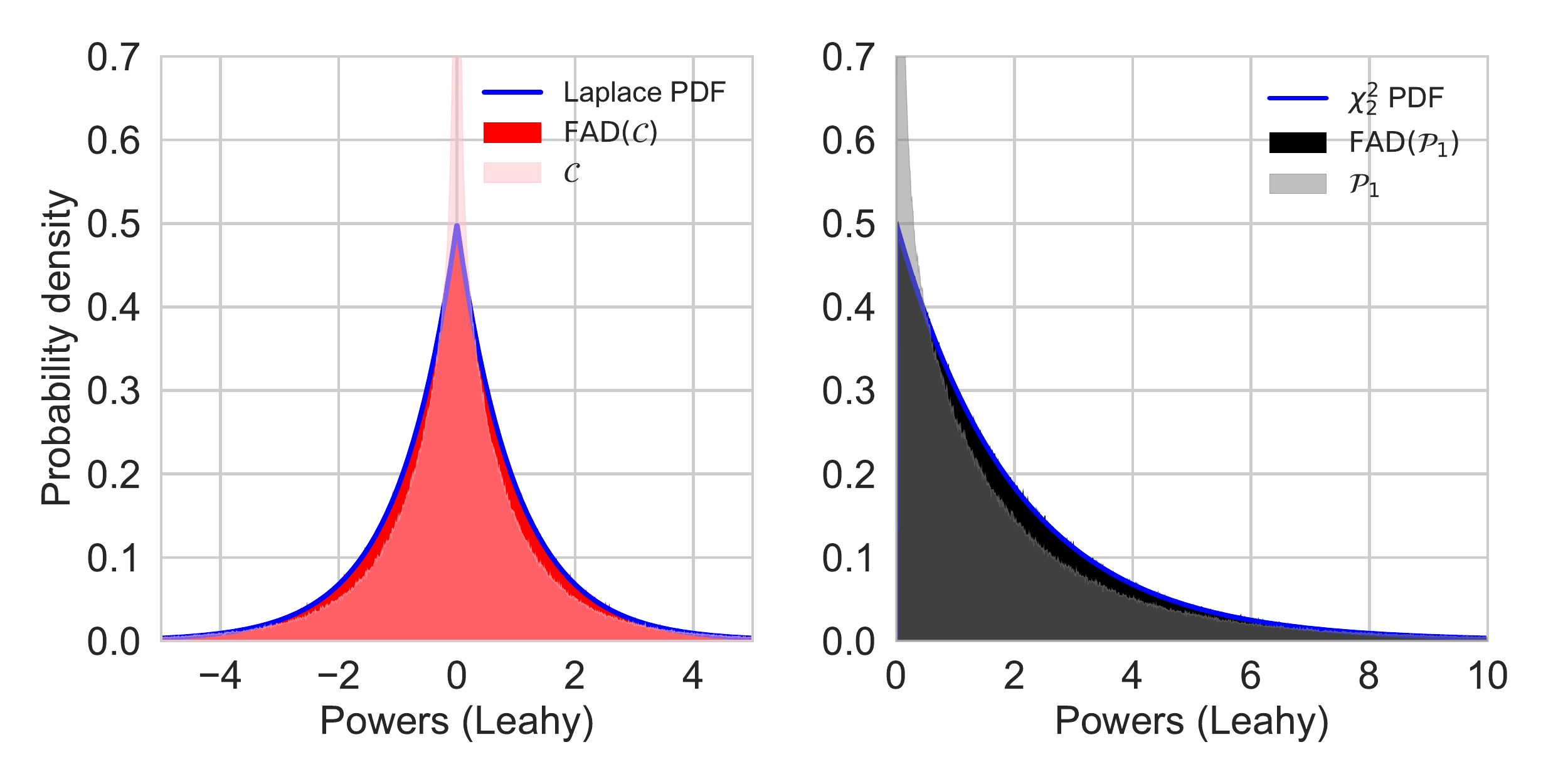}
\caption{Probability density function of non-averaged powers in the cospectrum (pink) and the periodogram (grey), before the FAD correction and after (red and black, respectively), shown as a fine-grained histogram. 
After correction, the powers follow remarkably well the expected Laplace (cospectrum) and $\chi^2_2$ (periodogram) distributions, as highlighted by the overplotted probability density functions (PDF).}
\label{fig:dist}
\end{figure*}

As laid out in Section \ref{sec:fourierdiff}, the difference of Fourier amplitudes from two independent but identical detectors shows no source variability, but \textit{still shows the same distortion} due to dead time (See Figure ~\ref{fig:fourierdiff}, left panel, where this is shown with red noise).
Let us simulate two constant 1000\,sec light curves with an incident mean count rate of 2000 counts/sec and a dead time of 2.5 ms, as we would expect from the two identical detectors of \nustar observing the same stable X-ray source.
The Fourier amplitudes of the light curves from the two detectors are heavily distorted by dead time, with the characteristic damped oscillator-like shape \citep{Vikhlinin+94,Zhang+95}  (Figure~\ref{fig:fourierdiff}, middle panel). 
Therefore, using the difference between the Fourier amplitudes in two detectors, we can in principle renormalize the periodogram so that only the source variability alters its otherwise flat shape.

As shown in Figure~\ref{fig:fourierdiff} (right panel), the single-detector Fourier amplitudes are proportional on average to the difference of the Fourier amplitudes in different realizations, with a constant factor $1/\sqrt{2}$.
Therefore, we expect that the periodogram will be proportional to the square of the Fourier amplitude difference, divided by 2.
Let us try to \textit{divide the periodogram by a smoothed version of the squared Fourier differences}, and multiply by 2.
For smoothing, we used a Gaussian running window with a window width of 50 bins.
Given that the initial binning had 50 bins/Hz, this interpolation allows an aggressive smoothing over bins whose y value does not change significantly.
We call this procedure the \textbf{Fourier Amplitude Difference} (hereafter FAD) \textbf{correction}.

The results of this correction are shown in Figure~\ref{fig:comparison}. 
Starting from a heavily distorted distribution of the powers, applying the FAD correction ``flattens'' remarkably well the white noise level of the periodogram and the distribution of the scatter of the white noise periodogram and cospectrum. 
Also, it reinstates a correct distribution of powers, following very closely the expected $\chi^2_2$ distribution (\citealt{Lewin+88}; Figure~\ref{fig:dist}, right). 
Analogously, the corrected cospectrum will follow the expected Laplace distribution (\citealt{HuppenkothenBachetti18}; Figure~\ref{fig:dist}, left).
While the original dead time-affected cospectrum had a distorted, frequency-dependent \rms level, the FAD-corrected cospectrum returns to the correct distribution at all frequencies.

%%%%%%%%%%%%%%%%%%%%%%%% SECTION 5 %%%%%%%%%%%%%%%%%%%%%%%%%%

\subsection{The FAD algorithm in detail} \label{sec:fad}
In practice, the FAD correction algorithm in a generic case would work as follows:
\begin{enumerate}
\item split the light curves from two independent, identical detectors into segments as one would do to calculate standard averaged periodograms;
\item for each pair of light curve segments: 
	\begin{itemize}
	\item calculate the Fourier transform of each detector separately, and then of the summed detectors (hereafter total-intensity); 
	\item save the unnormalized Fourier amplitudes;
	\item multiply these Fourier amplitudes by $\sqrt{2/N_{ph}}$ (that would give Leahy-normalized periodograms if squared);
	\item subtract the Leahy-normalized Fourier amplitudes of the two detectors between them, \textit{take the absolute value}, and obtain this way the Fourier Amplitude Differences (FAD);
	\item \textit{smooth} the FAD using a Gaussian-window interpolation with a large number of bins, in our case all the bins contained in 1-2 Hz, but this might need an adjustment at extreme count rates ($\gtrsim 10/\deadt$), where significant gradients in the white noise periodogram can occur in less than 1 Hz;
	\item use the separated single-detector and total-intensity \textit{unnormalized} Fourier amplitudes to calculate the periodograms (and/or the cospectrum);
	\item divide all periodograms (and/or the cospectrum) by the smoothed and squared FAD, and multiply by 2.
	\end{itemize}
\item normalize the periodograms to the wanted normalization (e.g. \citealt{Leahy+83} or fractional \rms: \citealt{BelloniHasinger90,Miyamoto+91}).
\end{enumerate}

%%%%%%%%%%%%%%%%%%%%%%%% SECTION 6 %%%%%%%%%%%%%%%%%%%%%%%%%%
\subsection{FAD correction of generic variable periodograms} \label{sec:correction}

\begin{figure*}
    \plotone{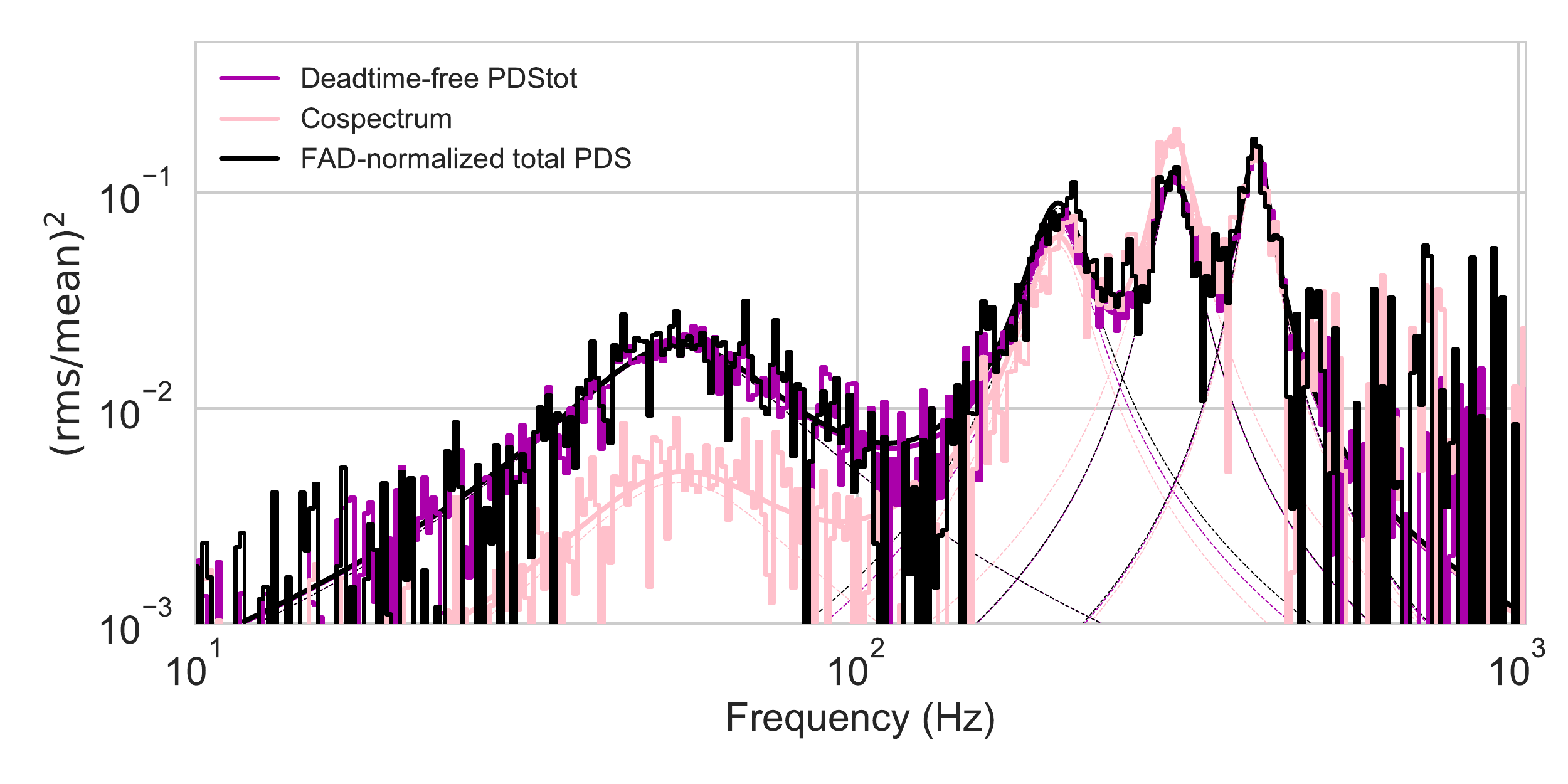}
    \caption{Periodogram, in fractional \rms normalization, from a simulation with four Lorentzian features (at 50, 200, 300 and 400 Hz) with 40-Hz full with at half maximum (FWHM). 
    We plotted and fitted the periodograms obtained before and after applying a 2.5\,ms dead time filter
    The total \rms before dead time was 30\% and the incident photon flux 400 ct/s. 
    There is no significant difference between the FAD-normalized and the dead-time-free periodogram, while the cospectrum without FAD (pink) clearly distorts the curves at different frequencies.
    }
    \label{fig:pds}
\end{figure*}

We are now ready to verify whether the FAD-corrected periodogram is a good approximation to the dead time-free periodogram.
To test this, we produced a number of different synthetic datasets as explained in Section~\ref{sec:wndeadtime}, containing different combinations of QPOs and broadband noise components, expressed as pairs of Lorentzian curves\footnote{See the notebooks at \href{https://github.com/matteobachetti/deadtime-paper-II}{https://github.com/matteobachetti/deadtime-paper-II} to reproduce the analysis plotted in the Figures of this paper and more. The algorithm described in Section~\ref{sec:fad} is contained in the \texttt{fad\_correction.py} file in the notebooks directory.}
We first calculated the periodogram and cospectrum of the dead time-free data, averaged over 64--512\,s intervals.
Then, we applied a 2.5\,ms dead time filter to the event list and applied the FAD correction, as described in Section~\ref{sec:fad}

All spectra were then expressed in fractional \rms normalization \citep{Miyamoto+91,BelloniHasinger90}.
In this normalization, the integrated model returns the full fractional \rms of the light curve, and the dead time-free and the FAD-corrected periodograms should be consistent over the full frequency range.
An example of this analysis is shown in Figure~\ref{fig:pds}: the FAD successfully corrects so well periodograms and cospectra when compared to dead-time free simulated spectra, that in the figure they are almost indistinguishable.

We ran extensive simulations testing how the method performs (1) for a range of different input count rates, leading to dead-time effects of different magnitude in the output periodograms and cospectra, and (2) when the light curves do not have the same count rate (since detectors may in reality have slightly different efficiencies).
We fitted all spectra with a two-Lorentzian model, plus a constant offset to account for the white noise level in periodograms.
We calculated the \rms by integrating the model fitted above over the full frequency range, and compared the results in the dead-time-free and FAD-corrected cases.

\subsection{Simulation results and discussion}\label{sec:caveat}

\begin{figure*}
    \plotone{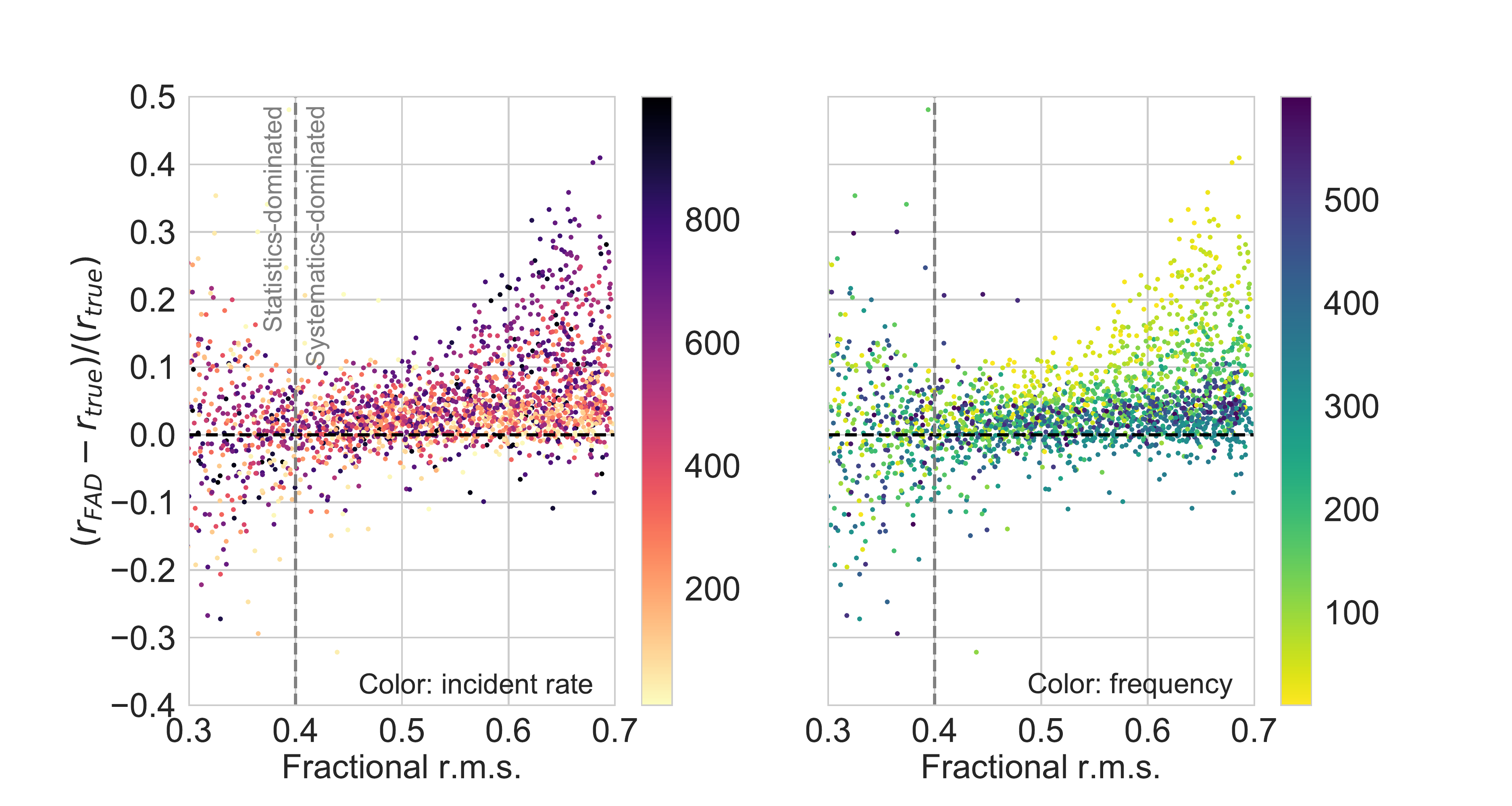}
    \caption{Relative overestimation of FAD with respect to \rms, versus \rms, as calculated from the cospectrum.  
    We encoded in the color the  the incident rate (left) and the frequency of the feature (right).
    From this visualization we see two regimes: below $\sim$40\% fractional \rms, the errors are dominated by statistical errors. 
    These errors will simply decrease when we average more data, as we expect from statistical errors.
    Over $\sim$40\% fractional \rms, FAD-corrected spectra overestimate the \rms, and this is 
    in particular when the incident rate is high, and the frequency relatively low. }
    \label{fig:errors}
\end{figure*}

The simulations described above show that the shape of the periodogram is precisely corrected by the FAD procedure if \textit{the input light curves have the same count rate} and \textit{for values of the input count rate and \rms that are not too extreme}. Differing input count rates in different detectors matter in practice only for the single-detector periodogram, but \textit{not} for cospectra and total-intensity periodograms. 
At high count rates, single-detector periodograms are corrected very well only if the two detectors have very similar count rates, and count rates must be more similar at higher count rates in order for the correction to apply.
However, we find that the total-intensity periodogram and the cospectrum remain well corrected by the FAD even if the count rate in the two detectors differs by 30\% in most cases.
Therefore, we recommend to use the FAD very carefully with single-detector periodograms, which should not be an issue given that the total-intensity periodogram is more sensitive and more convenient to use. 
A comparison with the cospectrum, which is not affected by white noise level distortions, is always recommended. 

However, we find that the FAD correction consistently overestimates the integrated \rms when the count rate and \rms are \textit{both} very high, \textit{in particular at low frequency} (See Figure~\ref{fig:errors}).
At $\sim$200 ct/s and 50\% \rms, the relative overestimation is below 5\% (meaning that if the true \rms is 50\%, the measured \rms is between 50 and 53\%) and it is symmetrically distributed around 0, as expected from statistical uncertainties. 
At higher incident rates and \rms, the uncertainty distribution is biased towards positive relative errors, implying an overestimation of the \rms. 
This should not be a problem in most use cases, when the \rms is used as a rough indicator for spectral state.
If very precise measurements of \rms are needed (for example, to calculate \rms/energy spectra), it is safer to account for this bias through simulations.
As a rough rule-of-thumb, the bias in the measured fractional \rms increases \textit{linearly} with the count rate and \textit{quadratically} with \rms.
A practical way to estimate this effect during analysis is to apply the FAD, obtain a best fit model, calculate the \rms, and simulate a number of realizations of the light curve to evaluate the amount of overestimation involved\footnote{relevant code can be found in Jupyter notebooks in the following GitHub repository: \href{https://github.com/matteobachetti/deadtime-paper-II}{https://github.com/matteobachetti/deadtime-paper-II}}.

%%%%%%%%%%%%%%%%%%%%%%%% SECTION 7 %%%%%%%%%%%%%%%%%%%%%%%%%%

\section{Conclusions}
In this Letter we described a method to correct the normalization of dead time-affected periodograms.
This method is valid in principle for 
1) correcting the shape of the periodogram, eliminating the well known pattern produced by dead time, and 
2) adjusting the white noise standard deviation of periodogram and cospectra to its correct value at all frequencies.
In general, we recommend applying the FAD correction to both the periodogram and the cospectrum. 
The periodogram, if obtained by the sum of the light curves, can yield a higher signal-to-noise ratio.
However, the white noise level subtraction is not always very precise due to mismatches in the mean count rate in the two light curves. 
A comparison with the FAD-corrected cospectrum, to verify visually the white noise subtraction, is always recommended: the white noise subtraction is the most important step when calculating the significance of a given feature in the periodogram \citep[e.g.][]{Barret+12,Huppenkothen+17}.
The cospectrum has indeed the advantage of not requiring white noise level subtraction.

We performed a number of simulations to test the validity of our method and explore its performance in the limits of high count rates as well as detectors with mis-matched efficiencies. 
In all cases, we find that the adjustment of the white noise standard deviation in the periodogram and the cospectrum works remarkably well, allowing to make a confident analysis of X-ray variability even in sources where this was precluded until now.
Only in cases where the count rate and the \rms are \textit{both} very high ($>$500 ct/s incident, $>$40\% resp.), the FAD correction leads to an overestimation the \rms, even if the white noise level of the periodogram remains flat.

\acknowledgments
We thank the anonymous referee for providing very insightful feedback.
We thank David W. Hogg for useful discussions on the topic of Fourier analysis, and Jeff Scargle for useful suggestions and comments.
MB is supported in part by the Italian Space Agency through agreement ASI-INAF n.2017-12-H.0 and ASI-INFN agreement n.2017-13-H.0.
DH is supported by the James Arthur Postdoctoral Fellowship and the Moore-Sloan Data Science Environment at New York University. DH acknowledges support from the DIRAC Institute in the Department of Astronomy at the University of Washington. The DIRAC Institute is supported through generous gifts from the Charles and Lisa Simonyi Fund for Arts and Sciences, and the Washington Research Foundation.

\software{Astropy \citep{astropy2013},
          Matplotlib \citep{Matplotlib07},
          scipy \& numpy \citep{numpy11},
          stingray \citep{huppenkothen2016},
          HENDRICS \citep[before name change]{2015ascl.soft02021B},
          Jupyter notebooks \citep{kluyver2016jupyter}}
\bibliographystyle{aasjournal}
%\bibliography{deadtime,papers3}

\begin{thebibliography}{}
\providecommand{\url}[1]{\href{#1}{#1}}

\bibitem[{{Astropy Collaboration} {et~al.}(2013){Astropy Collaboration},
  {Robitaille}, {Tollerud}, {Greenfield}, {Droettboom}, {Bray}, {Aldcroft},
  {Davis}, {Ginsburg}, {Price-Whelan}, {Kerzendorf}, {Conley}, {Crighton},
  {Barbary}, {Muna}, {Ferguson}, {Grollier}, {Parikh}, {Nair}, {Unther},
  {Deil}, {Woillez}, {Conseil}, {Kramer}, {Turner}, {Singer}, {Fox}, {Weaver},
  {Zabalza}, {Edwards}, {Azalee Bostroem}, {Burke}, {Casey}, {Crawford},
  {Dencheva}, {Ely}, {Jenness}, {Labrie}, {Lim}, {Pierfederici}, {Pontzen},
  {Ptak}, {Refsdal}, {Servillat}, \& {Streicher}}]{astropy2013}
{Astropy Collaboration}, {et~al.} 2013,
  \aap, 558, A33

\bibitem[{Astropy Project {et~al.}(2018)Price-Whelan, , G{\"u}nther, Lim,
  Crawford, Conseil, Shupe, Craig, Dencheva, Ginsburg, VanderPlas, Bradley,
  P{\'e}rez-Su{\'a}rez, de~Val-Borro, Aldcroft, Cruz, Robitaille, Tollerud,
  Ardelean, Babej, Bachetti, Bakanov, Bamford, Barentsen, Barmby, Baumbach,
  Berry, Biscani, Boquien, Bostroem, Bouma, Brammer, Bray, Breytenbach,
  Buddelmeijer, Burke, Calderone, Rodr{\'\i}guez, Cara, Cardoso, Cheedella,
  Copin, Crichton, D{\'A}vella, Deil, Depagne, Dietrich, Donath, Droettboom,
  Earl, Erben, Fabbro, Ferreira, Finethy, Fox, Garrison, Gibbons, Goldstein,
  Gommers, Greco, Greenfield, Groener, Grollier, Hagen, Hirst, Homeier, Horton,
  Hosseinzadeh, Hu, Hunkeler, , Jain, Jenness, Kanarek, Kendrew, Kern,
  Kerzendorf, Khvalko, King, Kirkby, Kulkarni, Kumar, Lee, Lenz, Littlefair,
  Ma, Macleod, Mastropietro, McCully, Montagnac, Morris, Mueller, Mumford,
  Muna, Murphy, Nelson, Nguyen, Ninan, N{\"o}the, Ogaz, Oh, Parejko, Parley,
  Pascual, Patil, Patil, Plunkett, Prochaska, Rastogi, Janga, Sabater,
  Sakurikar, Seifert, Sherbert, Sherwood-Taylor, Shih, Sick, Silbiger,
  Singanamalla, {Singer, L. P.}, Sladen, Sooley, Sornarajah, Streicher, Teuben,
  Thomas, Tremblay, Turner, Terr{\'o}n, van Kerkwijk, de~la Vega, Watkins,
  Weaver, Whitmore, Woillez, \& Zabalza}]{astropy18}
Astropy Project, {et~al.} 2018, arXiv,
  arXiv:1801.02634

\bibitem[{Bachetti(2015)}]{2015ascl.soft02021B}
Bachetti, M. 2015, MaLTPyNT: Quick look timing analysis for NuSTAR data, Astrophysics Source Code Library, ascl:1502.021

\bibitem[{Bachetti {et~al.}(2015)Bachetti, Harrison, Cook, Tomsick, Schmid,
  Grefenstette, Barret, Boggs, Christensen, Craig, Fabian, F{\"u}rst, Gandhi,
  Hailey, Kara, Maccarone, Miller, Pottschmidt, Stern, Uttley, Walton, Wilms,
  \& Zhang}]{Bachetti+15}
Bachetti, M., Harrison, F.~A., Cook, R., {et~al.} 2015, ApJ, 800, 109

\bibitem[{Barret \& Vaughan(2012)}]{Barret+12}
Barret, D., \& Vaughan, S. 2012, ApJ, 746, 131

\bibitem[{Belloni \& Hasinger(1990)}]{BelloniHasinger90}
Belloni, T., \& Hasinger, G. 1990, A{\&}A, 230, 103

\bibitem[{Hunter(2007)}]{Matplotlib07}
Hunter, J.~D. 2007, Computing in Science and Engineering, 9, 90

\bibitem[{Huppenkothen \& Bachetti(2017)}]{HuppenkothenBachetti18}
Huppenkothen, D., \& Bachetti, M. 2017, accepted to ApJ, arXiv:1709.09666

\bibitem[{{Huppenkothen} {et~al.}(2016){Huppenkothen}, {Bachetti}, {Stevens},
  {Migliari}, \& {Balm}}]{huppenkothen2016}
{Huppenkothen}, D., {Bachetti}, M., {Stevens}, A.~L., {Migliari}, S., \&
  {Balm}, P. 2016, {Stingray: Spectral-timing software}, Astrophysics Source
  Code Library, , , ascl:1608.001

\bibitem[{Huppenkothen {et~al.}(2017)Huppenkothen, Younes, Ingram, Kouveliotou,
  G{\"o}{\u{g}}{\"u}{\c s}, Bachetti, Sanchez-Fernandez, Chenevez, Motta,
  van~der Klis, Granot, Gehrels, Kuulkers, Tomsick, \&
  Walton}]{Huppenkothen+17}
Huppenkothen, D., Younes, G., Ingram, A., {et~al.} 2017, ApJ, 834, 90

\bibitem[{Kluyver {et~al.}(2016)Kluyver, Ragan-Kelley, P{\'e}rez, Granger,
  Bussonnier, Frederic, Kelley, Hamrick, Grout, Corlay,
  {et~al.}}]{kluyver2016jupyter}
Kluyver, T., Ragan-Kelley, B., P{\'e}rez, F., {et~al.} 2016, in ``Positioning and Power in Academic Publishing: Players, Agents and Agendas'', 87--90

\bibitem[{Leahy {et~al.}(1983)Leahy, Darbro, Elsner, Weisskopf, Kahn,
  Sutherland, \& Grindlay}]{Leahy+83}
Leahy, D.~A., Darbro, W., Elsner, R.~F., {et~al.} 1983, ApJ, 266, 160

\bibitem[{Lewin {et~al.}(1988)Lewin, van Paradijs, \& van~der Klis}]{Lewin+88}
Lewin, W. H.~G., van Paradijs, J., \& van~der Klis, M. 1988, SSRv, 46, 273

\bibitem[{Miyamoto {et~al.}(1991)Miyamoto, Kimura, Kitamoto, Dotani, \&
  Ebisawa}]{Miyamoto+91}
Miyamoto, S., Kimura, K., Kitamoto, S., Dotani, T., \& Ebisawa, K. 1991, ApJ,
  383, 784

\bibitem[{{Timmer} \& {Koenig}(1995)}]{timmer1995}
{Timmer}, J., \& {Koenig}, M. 1995, \aap, 300, 707

\bibitem[{van~der Walt {et~al.}(2011)van~der Walt, Colbert, \&
  Varoquaux}]{numpy11}
van~der Walt, S., Colbert, S.~C., \& Varoquaux, G. 2011, Computing in Science
  Engineering, 13, 22

\bibitem[{Vikhlinin {et~al.}(1994)Vikhlinin, Churazov, \&
  Gilfanov}]{Vikhlinin+94}
Vikhlinin, A., Churazov, E., \& Gilfanov, M. 1994, A\&A, 287, 73

\bibitem[{Zhang {et~al.}(1995)Zhang, Jahoda, Swank, Morgan, \&
  Giles}]{Zhang+95}
Zhang, W., Jahoda, K., Swank, J.~H., Morgan, E.~H., \& Giles, A.~B. 1995, ApJ,
  449, 930

\end{thebibliography}

\end{document}